\documentstyle[preprint,aps]{revtex}
\tighten
\begin{document}
\title{Gauss's Law, Gauge-Invariant States, and
 Spin and Statistics In Abelian Chern-Simons Theories.}
\author{Kurt Haller\thanks{ e-mail address: KHALLER@UCONNVM.UCONN.EDU}
 and Edwin Lim-Lombridas\thanks{present address: Guardian Life
Insurance Co., New York, NY 10003}}
\address{Department of Physics, University of Connecticut, Storrs, Connecticut
06269}
\maketitle
\begin{abstract}

We discuss topologically massive QED\,---\,the Abelian gauge theory in
which (2+1)-dimensional QED with a Chern-Simons term is minimally coupled to a
spinor field.  We quantize the theory in covariant gauges, and construct a
class of
unitary transformations that enable us to embed the theory in a Fock space
of states
that implement Gauss's law.  We show that when electron (and positron)
creation and
annihilation operators represent gauge-invariant charged particles that are
surrounded
by the electric and magnetic fields required by Gauss's law, the unitarity
of the
theory is manifest, and charged particles interact with photons and with
each other
through nonlocal potentials.  These potentials include a Hopf-like
interaction, and a
planar analog of the Coulomb interaction.  The gauge-invariant charged particle
excitations that implement Gauss's law obey the identical anticommutation
rules as do
the original gauge-dependent ones. Rotational phases, commonly
identified as planar `spin', are arbitrary, however.

\end{abstract}

\section{Introduction}
\noindent In this report we will summarize and extend our earlier work on
Maxwell-Chern-Simons  (MCS) and Chern-Simons (CS)
theory,\cite{cs,mcstemp,mcscov} and discuss the properties of the particle
excitations produced when relativistic fields\,---\,spinors, in this
case\,---\,are
minimally coupled to Abelian (2+1)-dimensional gauge theories that include
Chern-Simons  terms.  This work will also illustrate canonical quantization as a
method for exploring  the properties of the particle excitations of gauge fields
and of the fields coupled to  them.  In this, as in earlier work\,---\,we will
use MCS theory in covariant gauges as  an illustrative example
here\,---\,we will
construct gauge-invariant  excitations of the charged spinor fields  that
satisfy Gauss's law and the appropriate gauge condition.  We will also
demonstrate  that these charged particle excitations\,---\,we will refer to them
as `electrons' and `positrons,' even though  they appear in a (2+1)-dimensional
space\,---\,interact through an effective Hamiltonian that  consists of a vertex
interaction at which two charged spinor lines meet a topologically  massive
`photon,' and through nonlocal interactions, one of which has a long-range
tail with an Aharonov-Bohm-like structure (the so-called Hopf
interaction).\cite{forte} One of our main  objectives is to explore the
rotational and statistical properties of the gauge-invariant particle states
that obey Gauss's law.

\section{Dynamical Considerations}
\noindent The Lagrangian for MCS theory minimally coupled to a spinor field is
\begin{equation} {\cal L}_{\mbox{\scriptsize MCS}} =
-\frac{1}{4}\,F_{\mu\nu}F^{\mu\nu} +
\frac{1}{4}\,m\epsilon_{\mu\nu\lambda}F^{\mu\nu}A^\lambda-j_\mu A^\mu -
\bar{\psi}(M - i\gamma_\mu\partial^\mu)\psi + {\cal L}_{\mbox{\scriptsize
gf}}
\end{equation}
where $j^\mu = e\bar{\psi}\gamma^\mu\psi$ and $m$ is the
Chern-Simons coefficient which,  in Abelian theories,  can have arbitrary real
values. ${\cal L}_{\mbox{\scriptsize gf}}$  represents the gauge-fixing
Lagrangian; in our work, we have used forms that correspond to  the following
gauges:
\begin{equation} {\cal L}_{\mbox{\scriptsize gf}} = \left\{
\begin{array}{ll} -G\partial_\mu A^\mu - \frac{1}{2}\,(1-\gamma)G^2 & \ \ \
\mbox{\small covariant} \\[10pt]
\ \ \ \ \ \ \ \ \gamma=0 \rightarrow \mbox{\small Feynman}\\
\ \ \ \ \ \ \ \ \gamma=1 \rightarrow \mbox{\small Landau}\\[6pt] -G\partial_0A_0
& \ \ \ \mbox{\small temporal}\\[6pt] -G\nabla\cdot{\bf A} & \ \ \ \mbox{\small
Coulomb}.
\end{array}
\right.
\end{equation}
In covariant gauges, there are trivial constraints only, so that
the canonical quantization of this model can be carried out without
requiring any
special procedures for implementing primary constraints.     All components
of the
gauge field,
$A_0$ as well as $A_l,$ have canonically conjugate  momenta given by
\begin{equation}
\Pi_l = F_{0l} + \frac{1}{2}\,m\epsilon_{ln}A_n\ \ \ (F_{0l} =
-E_l)\ \ \mbox{and}\ \ \Pi_0 = -G,
\end{equation}
and they are subject to the standard equal-time commutation or
anticommutation rules
(ETCR)
\begin{equation} [A_0({\bf x}),G({\bf y})] = -i\delta({\bf x-y}),
\label{eq:AcomG}
\end{equation}
\begin{equation} [A_l({\bf x}),\Pi_n({\bf y})] = i\delta_{ln}\delta({\bf x-y}),
\label{eq:AcomPi}
\end{equation}
\begin{equation}
\mbox{and}\;\;\;\;\;\;\;\;\;\{\psi({\bf x}),\psi^\dagger({\bf y})\} =
\delta({\bf x-y}).
\end{equation} The Hamiltonian for covariant gauges is given by $H = H_0 +
H_{\mbox{\scriptsize I}},$ where
\begin{eqnarray} H_0 &=& \int d{\bf x}\ \left[\frac{1}{2}\,\Pi_l\Pi_l +
\frac{1}{4}\,F_{ln}F_{ln} +
G\partial_lA_l-\frac{1}{2}\,(1-\gamma)G^2 + \frac{1}{8}\,m^2
A_lA_l\right.\nonumber\\
&&\ \ \ \ +\ \left.A_0\left(\partial_l\Pi_l -
\frac{1}{4}\,m\epsilon_{ln}F_{ln}+j_0\right) -j_lA_l
+\frac{1}{2}\,m\epsilon_{ln}A_l\Pi_n\right] + H_{e\bar{e}}
\end{eqnarray}
and
\begin{equation}
H_{\mbox{\scriptsize I}} = \int d{\bf x}\ \left(j_0A_0
- j_lA_l\right).
\end{equation} In order to associate these fields with particle excitations, we
expand the operator-valued  fields in terms of creation and annihilation
operators for particle excitations. These  excitations include two excitation
modes of the spinor field $\psi({\bf x})$ --- `electrons' with  spin $1/2$ and
`positrons' with spin $-1/2$ --- and a single propagating topologically
massive  mode
of the gauge fields, originally described by Deser, Jackiw and
Templeton.\cite{djt}  Annihilation  operators for electrons and positrons of
momentum ${\bf k}$ are represented as $e_{\scriptsize k}$ and
$\bar{e}_{\bf k}$ respectively, and the corresponding creation operators are
$e^\dagger_{\bf k}$ and
$\bar{e}^\dagger_{\bf k}.$ Annihilation and creation operators for the
topologically massive photons  are represented as $a({\bf k})$ and
$a^\dagger({\bf k})$ respectively. In the covariant gauge, and in any
gauge with only trivial primary constraints,  `ghost' modes are also necessary
to represent the ETCR for all components of the gauge field. There are two
different varieties of ghost modes, `Q' and `R' ghosts, whose annihilation and
creation operators, respectively, are
 $a_Q({\bf k}), a_R({\bf k}),$ and $a_Q^{\mbox{\normalsize$\star$}}({\bf k})$,
$a_R^{\mbox{\normalsize$\star$}}({\bf k}).$ Since the one-particle ghost states
must have zero norm,
$a_Q({\bf k})$ and its conjugate creation operator
$a_Q^{\mbox{\normalsize$\star$}}({\bf k})$ must commute, as must
$a_R({\bf k})$ and $a_R^{\mbox{\normalsize$\star$}}({\bf k}).$  But since the
ghost modes must support the  ETCR of the gauge fields, not {\em all} the ghost
operators can commute. The commutation rules for  ghost operators are
\begin{equation} [a_Q({\bf k}),a_R^{\mbox{\normalsize $\star$}}({\bf q})] =
[a_R({\bf k}),a_Q^{\mbox{\normalsize $\star$}}({\bf q})] = \delta_{\bf kq},
\end{equation} and the unit operator in the one particle ghost (OPG) sector is
\begin{equation} 1_{\mbox{\scriptsize opg}} = \sum_{\bf
k}\left[a_Q^{\mbox{\normalsize $\star$}}({\bf k})|0\rangle\langle 0|a_R({\bf k})
+ a_R^{\mbox{\normalsize $\star$}}({\bf k})|0\rangle\langle 0|a_Q({\bf
k})\right].
\end{equation}

The excitation operators we have described are used to represent the spinor and
gauge fields in such a  way that they implement the space-time commutation
rules, Eqs.~(\ref{eq:AcomG})--(\ref{eq:AcomPi}).  They must also result in a
Hamiltonian\,---\,the `free field' Hamiltonian, in the first
instance\,---\,that
produces only one kind of ghost in the course of time evolution.  When only
a single
variety of ghost  coexists with the propagating observable states (we will
always
select the `Q' ghost to fulfill this role),  then the states with ghost
content are
purely zero-norm states, and they do not drain any probability from  the
part of the
Hilbert space that represents physically observable configurations.  When
the two
ghosts  coexist, the states with mixed ghost content are not zero-norm
states, and
threaten the unitarity of the  theory by draining probability from the space of
propagating particles to the `ghost' part of the Hilbert space.

The representation of the gauge fields in terms of the particle and ghost
annihilation and creation operators  is given in Ref.~[3],
and lack of space prevents its repetition here. But we will specify the  `free'
Hamiltonian,
$H_0,$ which is given by
\begin{eqnarray} H_0 &=& \sum_{\bf k}\frac{\omega_k}{2}\left[a^\dagger({\bf
k})a({\bf k}) + a({\bf k})a^\dagger({\bf k})\right]+ \sum_{\bf
k}k\left[a_Q^{\mbox{\normalsize $\star$}}({\bf k})a_R({\bf k}) + a_Q({\bf
k})a_R^{\mbox{\normalsize $\star$}}({\bf k})\right]\nonumber\\
&-&(1-\gamma)\sum_{\bf k}\frac{64k^4}{m^3}\,a_Q^{\mbox{\normalsize
$\star$}}({\bf k})a_Q({\bf k}) + \int d{\bf x}\ \psi^\dagger(\gamma_0M -
i\gamma_0\gamma_l\partial_l)\psi.
\label{eq:Hfree}
\end{eqnarray}  The interaction Hamiltonian, $H_{\mbox{\scriptsize I}},$ in the
same representation, is also given in Ref.~[3].  In contrast to $H_0,$
$H_{\mbox{\scriptsize I}}$ has a profusion of {\em both}~varieties of ghost
creation and annihilation operators, so that we must  take further steps to
protect the unitarity of the theory.  We also have to respond to the fact that
$e_{\bf k}$ and $\bar{e}_{\bf k}$ annihilate, and  $e^\dagger_{\bf k}$ and
$\bar{e}^\dagger_{\bf k}$  create, electrons and positrons that are entirely
`free'\,---\,$i.e.$ unaccompanied by any electric  and magnetic fields\,---\,so
that they fail to implement Gauss's law.

\section{Implementing Gauss's Law}
\noindent Gauss's law for MCS theory is expressed by
\begin{equation}
\partial_lF_{0l} -\frac{1}{2}\,m\epsilon_{ln}F_{ln} + j_0=0
\label{eq:Gauss}
\end{equation} and the `Gauss's law operator' ${\cal G}({\bf x})$ is given by
\begin{equation}
{\cal G}= \partial_lF_{0l}
-\frac{1}{2}\,m\epsilon_{ln}F_{ln} + j_0= \partial_l\Pi_l -
\frac{1}{4}\,m\epsilon_{ln}F_{ln} + j_0.
\label{eq:Gausslaw}
\end{equation}
${\cal G}({\bf x})$ can be represented as
\begin{equation} {\cal G}({\bf x}) = \sum_{\bf
k}\frac{8k^3}{m^{3/2}}\left[\Omega({\bf k})e^{i{\bf k\cdot
x}}+\Omega^{\mbox{\normalsize $\star$}}({\bf k})e^{-i{\bf k\cdot x}}\right],
\end{equation} where
\begin{equation}
\Omega({\bf k}) = a_Q({\bf k}) + \frac{m^{3/2}}{16k^3}\,j_0({\bf k})
\end{equation} with
\begin{equation} j_0({\bf k})=\int d{\bf x}\ j_0({\bf x})e^{-i{\bf k\cdot x}}.
\end{equation}
$\Omega({\bf k})$ can be used to impose Gauss's law on a set of states
$\{|\nu\rangle\}$ since it obeys
\begin{equation} [H, \Omega({\bf k})] = -k\Omega({\bf k})
\end{equation} so that $\Omega({\bf k})$ is operator-valued, but has a
$c$-number
time dependence in the Heisenberg picture,
\begin{equation}
\Omega({\bf k},t) = e^{iHt}\Omega({\bf k})e^{-iHt}=\Omega({\bf k})e^{-ikt}.
\end{equation}
Hence,
\begin{equation}
\Omega({\bf k})|\nu\rangle = 0,
\label{eq:subcon}
\end{equation} if imposed at one time $(t=0)$ remains in force forever!

We now observe that, because the gauge group of this theory is Abelian,
$\Omega({\bf k})$ is unitarily equivalent  to $a_Q({\bf k}),$ so that there are
unitary operators, $U,$\,---\,many, in fact\,---\,for which
\begin{equation} U^{-1}\Omega({\bf k})U = a_Q({\bf k}).
\end{equation}
 Examples of $U$ operators are given in Ref.~[3], but, because of
limited space,  will not be repeated here.  The important point to emphasize is
that there are two alternate  ways of using $U$ operators to construct
states that
implement Gauss's law: We can use
$U$ to construct states $|\nu\rangle$ that satisfy
$\Omega({\bf k})|\nu\rangle=0$, for which $H$ is the generator of
time-evolutions.  Alternatively\,---\,and  this is the option we will use
here\,---\,we
can transpose the {\em entire formulation}  to a new representation using
\begin{equation}
\hat{\cal P} = U^{-1}{\cal P}U\ \ \ \mbox{\small (${\cal P}$ represents any
operator)}
\end{equation} to transform all operators to a new representation.   Then
$U^{-1}\Omega({\bf k})U = \hat{\Omega}({\bf k}) = a_Q({\bf k})$,
\begin{equation} U^{-1}HU = \hat{H}
\end{equation} and similarly for all gauge field (and spinor) operators, as well
as all dynamical variables. This similarity transformation preserves all
algebraic relationships among transformed operators.   The unitary transforms of
the states, $|\nu\rangle$, that satisfy Eq.~(\ref{eq:subcon}) are given by
\begin{equation}
\Omega({\bf k})|\nu\rangle = 0\ \ \Rightarrow\ \ \hat{\Omega}({\bf k})|n\rangle
= a_Q({\bf k})|n\rangle = 0.
\label{eq:subcontrans}
\end{equation} The states, $|n\rangle$, constitute the Fock space built on the
perturbative vacuum $|0\rangle$  annihilated by $e({\bf q})$, $\bar{e}({\bf
q})$, $a({\bf q})$,
$a_Q({\bf q})$, $a_R({\bf q})$, in which creation operators that commute with
$a_Q({\bf k})$  act on $|0\rangle$\,---\,for example, {\em $e^\dagger({\bf
q})|0\rangle$,
$\bar{e}^\dagger({\bf q})|0\rangle$, $a^\dagger({\bf q})|0\rangle$,
 $a_Q^{\mbox{\normalsize $\star$}}({\bf q})|0\rangle$, {\em  but not}
$a_R^{\mbox{\normalsize $\star$}}({\bf q})|0\rangle$}.
 Moreover, in the transformed $\hat{}$ representation, because
Eq.~(\ref{eq:subcontrans}) imposes  Gauss's law on the transformed states,
{$|e({\bf q})\rangle = e^\dagger({\bf q})|0\rangle$ represents an electron {\em
with} the electric  and magnetic fields  required to obey Gauss's law.} The
unitary transforms of the gauge fields are readily evaluated, but only the most
interesting\,---\,$\hat{E}_l({\bf x})$ and $\hat{B}({\bf x})$\,---\,will be
given here.
They are:
\begin{equation}
\hat{E}_l({\bf x}) = E_l({\bf x}) -\frac{1}{2\pi}\,\frac{\partial}{\partial
x_l}\,\int d{\bf y}\ K_0(m|{\bf x-y}|)j_0({\bf y})
\label{eq:efield}
\end{equation} and
\begin{equation}
\hat{B}({\bf x}) = B({\bf x}) -\frac{m}{2\pi}\,\int d{\bf y}\ K_0(m|{\bf
x-y}|)j_0({\bf y})
\label{eq:bfield}
\end{equation} where $E_l({\bf x})$ and $B({\bf x})$ are the untransformed
fields, and  include only the contributions from the topologically massive
photons  and the ghost components of $\hat{E}_l({\bf x})$ and $\hat{B}({\bf
x}).$ They are:
\begin{eqnarray} E_l({\bf x}) &=& -\sum_{\bf
k}\frac{imk_l}{k\sqrt{2\omega_k}}\left[a({\bf k})e^{i{\bf k\cdot
x}}-a^\dagger({\bf k})e^{-i{\bf k\cdot x}}\right]\nonumber\\ &-&\sum_{\bf
k}\frac{\sqrt{\omega_k}\epsilon_{ln}k_n}{\sqrt{2}k}\left[a({\bf k})e^{i{\bf
k\cdot x}}+a^\dagger({\bf k})e^{-i{\bf k\cdot x}}\right]\nonumber\\ &-&\sum_{\bf
k}\frac{8k^2\epsilon_{ln}k_n}{m^{5/2}}\left[a_Q({\bf k})e^{i{\bf k\cdot
x}}+a_Q^{\mbox{\normalsize $\star$}}({\bf k})e^{-i{\bf k\cdot x}}\right]
\end{eqnarray} and
\begin{eqnarray} B({\bf x}) &=& \sum_{\bf
k}\frac{k}{\sqrt{2\omega_k}}\left[a({\bf k})e^{i{\bf k\cdot x}}+a^\dagger({\bf
k})e^{-i{\bf k\cdot x}}\right]\nonumber\\ &+& \sum_{\bf
k}\frac{8k^3}{m^{5/2}}\left[a_Q({\bf k})e^{i{\bf k\cdot
x}}+a_Q^{\mbox{\normalsize $\star$}}({\bf k})e^{-i{\bf k\cdot x}}\right].
\end{eqnarray}
Equations~(\ref{eq:efield}) and (\ref{eq:bfield}) show that
combinations of magnetic fields and  longitudinal electric fields accompany
charge densities in the transformed $\,{\bf \hat{}}\,$ representation,
implementing Gauss's law for the Fock space of charged states that solve
Eq.~(\ref{eq:subcontrans}). The transformed Hamiltonian $\hat{H}$ is given by
\begin{equation}
\hat{H} = \hat{H}_{\mbox{\scriptsize quot}} + h + H_Q
\end{equation} where
$h$ has no dynamical significance because it is a total time derivative\,---\,it
can be expresses as
 $h = i[H_0,\chi]$ or, equivalently, as  $h = i[\hat{H},\chi]$ where
\begin{equation}
\chi = -\sum_{\bf k}\frac{3m^{3/2}\phi({\bf k})}{32k^3}\,j_0({\bf k})j_0(-{\bf
k}).
\end{equation}
$H_Q$ is complicated, but entirely composed of parts that are proportional to
$a_Q$ and $a_Q^{\mbox{\normalsize $\star$}}$; thus, it plays {\em no role} in
dynamical  time-evolution of state vectors.
$\hat{H}_{\mbox{\scriptsize quot}}$ operates within a quotient space of charged
excitations (the `electrons' and  `positrons') and the topologically massive
photons.  It has the form
\begin{equation}
\hat{H}_{\mbox{\scriptsize quot}} = H_{e\bar{e}}+ \sum_{\bf
k}\frac{\omega_k}{2}\left[a^\dagger({\bf k})a({\bf k}) + a({\bf
k})a^\dagger({\bf k})\right] + \hat{H}_{\mbox{\scriptsize I}}
\end{equation} where
$H_{e\bar{e}}$ `counts' the  topologically massive photons  and electrons
(positrons) and assigns them their appropriate energies.
 $\hat{H}_{\mbox{\scriptsize I}}$ is given by
\begin{equation}
\hat{H}_{\mbox{\scriptsize I}}=H_{\mbox{\scriptsize a}}+H_{\mbox{\scriptsize
b}}+H_{j\gamma}
\end{equation} where $H_{\mbox{\scriptsize b}}$ is an interaction between charge
density and transverse current  densities,
\begin{equation} H_{\mbox{\scriptsize b}} = \int d{\bf x}\,d{\bf y}\ j_0({\bf
x})\epsilon_{ln}j_l({\bf y})(x-y)_n{\cal F}(|{\bf x-y}|)
\end{equation} and ${\cal F}(R)$ is a nonlocal interaction given by
\begin{equation} {\cal F}(R) = -\frac{m}{2\pi}\int_0^\infty du\
\frac{J_1(u)}{u^2+(mR)^2} \rightarrow \left\{\begin{array}{ll} 1/4\pi R & \ \ mR
\rightarrow 0\\[12pt] 1/2\pi m R^2 & \ \ mR \rightarrow
\infty\end{array}\right..
\end{equation}
 $H_{\mbox{\scriptsize a}}$ is a planar analog of the Coulomb interaction
\begin{equation} H_{\mbox{\scriptsize a}} = \int d{\bf x}\,d{\bf y}\ j_0({\bf
x})j_0({\bf y})K_0(m|{\bf x-y}|)
\end{equation} where $K_0(m|{\bf x-y}|)$ is the modified Bessel function;
$K_0(u) \rightarrow \infty$ logarithmically  when $u\rightarrow 0$.
$H_{\mbox{\scriptsize a}}$ is defined, however, for well-behaved $j_0({\bf x})$
because the integration  compensates for the singularity in  $K_0(m|{\bf
x-y}|)$.
$H_{j\gamma}$ is an interaction between topologically massive photons and
currents, given by:
\begin{eqnarray} H_{j\gamma} &=&\sum_{\bf
k}\frac{mk_l}{\sqrt{2}k\omega_k^{3/2}}\left[a({\bf k})j_l(-{\bf k}) +
a^\dagger({\bf k})j_l({\bf k})\right]\nonumber\\
&-&\sum_{\bf
k}\frac{i\epsilon_{ln}k_n}{k\sqrt{2\omega_k}}\left[a({\bf k})j_l(-{\bf k}) -
a^\dagger({\bf k})j_l({\bf k})\right].
\end{eqnarray}
$\hat{H},$ which operates in a Fock space of states for which Gauss's law has
been implemented, incorporates the  feature that it {\em never} time-evolves
states from the space of physical, propagating particles  to the part of Hilbert
space in which different varieties of ghosts coexist. The implementation of
Gauss's law  therefore naturally protects the unitarity of the theory.   It is
of special interest that $\hat{H}_{\mbox{\scriptsize quot}}$ is identical in
{\em all} gauges  (temporal, covariant, Coulomb, etc.) when Hamiltonians are
unitarily transformed to the ``standard''  representation in which
Eqs.~(\ref{eq:efield}) and (\ref{eq:bfield}) describe the relation between
charges and  the fields that surround them.
 Only the physically irrelevant $H_Q$ depends on the gauge (for example, $H_Q =
0$ in the Coulomb gauge.)

\section{Gauge-Invariance and Gauss's Law}
\noindent The Gauss's law operator ${\cal G}$ is the generator of gauge
transformations:
\begin{equation} e^{-i\int d{\bf y}\ {\cal G}({\bf y})\chi({\bf
y})}\left\{\begin{array}{c} A_i({\bf x}) \\ \psi({\bf
x})\end{array}\right\}e^{i\int d{\bf y}\ {\cal G}({\bf y})\chi({\bf y})}
= \left\{\begin{array}{c} A_i^\prime = A_i
+\partial_i\chi\\\nonumber
\psi^\prime = e^{ie\chi}\psi\end{array}\right\}.
\end{equation} In the `transformed' $\,\hat{}\,$ representation,
\begin{equation} \hat{\cal G} = \partial_l\hat{\Pi}_l -
\frac{1}{4}\,m\epsilon_{ln}\hat{F}_{ln} + j_0=\partial_l\Pi_l -
\frac{1}{4}\,m\epsilon_{ln}F_{ln}
\end{equation} ($j_0$ is absent in the `transformed' $\,\hat{}\,$
representation, although it is implicit in $\Pi_l$ and $A_l$).
$\hat{\cal G}$ manifestly commutes with $\psi,$ so that $\psi$ is a
gauge-invariant spinor field in  the `transformed' $\,\hat{}\,$ representation:
\begin{equation}
\psi = \hat{\psi}_{\mbox{\scriptsize GI}}
\end{equation}

What is $\hat{\psi}_{\mbox{\scriptsize GI}}$ in the original untransformed
representation? We can use the $U$ operator to transform {\em back} from the
`transformed' $\,\hat{}\,$ representation to the `original' untransformed
representation as shown by\footnote{This corrects an error in the
specification of the
gauge-invariant spinor operators in our Refs. [2] and [3].}
\begin{equation} {\psi}_{\mbox{\scriptsize GI}} = U\hat{\psi}_{\mbox{\scriptsize
GI}}U^{-1} =  U\psi U^{-1} = e^{{\cal D}^\prime(x)}\psi(x),
\end{equation} where
\begin{eqnarray} {\cal D}^\prime({\bf x}) &=& -ie\int d{\bf y}\
\left\{\rule{0pt}{16pt}\xi_1^\prime(|{\bf
x-y}|)\epsilon_{ln}\partial_l\Pi_n({\bf y}) + \xi_2^\prime(|{\bf
x-y}|)\partial_lA_l({\bf y})\right.\nonumber\\
&+&\eta_1^\prime(|{\bf
x-y}|)\partial_l\Pi_l({\bf x})+\eta_2^\prime(|{\bf
x-y}|)\epsilon_{ln}\partial_lA_n({\bf x})\nonumber\\
&+&\left.\left[\chi^\prime(|{\bf x-y}|)+
\zeta^\prime(|{\bf x-y}|)\right]G({\bf x})\rule{0pt}{16pt}\right\}
\end{eqnarray} and where
\begin{equation}
\xi_1^\prime(|{\bf x-y}|) = \sum_{\bf k}\frac{m}{\omega_k^2k^2}\,e^{i{\bf
k\cdot(x-y)}},
\end{equation}
\begin{equation}
\xi_2^\prime(|{\bf x-y}|) = -\sum_{\bf k}\frac{1}{2}\left(\frac{1}{\omega_k^2} +
\frac{1}{k^2}\right)e^{i{\bf k\cdot(x-y)}},
\end{equation}
\begin{equation}
\eta_1^\prime(|{\bf x-y}|) = \sum_{\bf k}\frac{m^{3/2}\phi({\bf
k})}{8k^3}\,e^{i{\bf k\cdot(x-y)}},
\end{equation}
\begin{equation}
\eta_2^\prime(|{\bf x-y}|) = \sum_{\bf k}\left(\frac{m^{5/2}\phi({\bf
k})}{16k^3} - \frac{1}{m\omega_k^2}\right)e^{i{\bf k\cdot(x-y)}},
\end{equation}
\begin{equation}
\chi^\prime(|{\bf x-y}|) = \sum_{\bf k}
\frac{1}{4k^2}\left[(1-\gamma)+\frac{2k^2}{m^2}-
\frac{k^4}{m^2\omega^2}\right]e^{i{\bf k\cdot(x-y)}}\nonumber\\
\end{equation} and
\begin{equation}
\zeta^\prime(|{\bf x-y}|) = -\sum_{\bf k}\frac{m^{3/2}\theta({\bf
k})}{8k^2}\,e^{i{\bf k\cdot(x-y)}}.
\end{equation}
$\theta({\bf k})$ and $\phi({\bf k})$ are optional constituents of ${\cal
D}^\prime({\bf x})$. They  may be set to zero, or they may be any arbitrary
real and
even functions of ${\bf k}$.

In ${\cal D}^\prime(x)$, if the gauge fields are gauge-transformed,
\begin{equation} A_l \rightarrow A_l + \partial_l\chi,
\end{equation}
\begin{equation}
\Pi_l \rightarrow \Pi_l + \frac{1}{2}\,m\epsilon_{ln}\partial_n\chi,
\end{equation} and, consequently,
\begin{equation} {\cal D}^\prime(x) \rightarrow {\cal D}^\prime(x) - ie\chi
\end{equation}

Since $\psi\rightarrow\psi e^{ie\chi}$, $e^{{\cal D}^\prime(x)}\psi(x)$ is
gauge-invariant. Also, $e^{{\cal D}^\prime(x)}\psi(x)$ creates particles from
the vacuum that obey Gauss's law.
\begin{equation} [{\cal G}({y}), e^{{\cal D}^\prime(x)}\psi(x)] = 0.
\end{equation} This confirms our earlier demonstration that the charged
particles, whose physical interactions are described by
$\hat{H}_{\mbox{\scriptsize I}},$ are gauge-invariant states and obey Gauss's
law.

\section{Rotations, Particle Exchanges, Statistics, etc.}

\noindent
{\underline{Statistical properties are unaffected by the
implementation of Gauss's law:} \\
\noindent Since unitary transformations  do not change algebraic relations,
such as
equal-time commutation and  anticommutation rules, we observe that
\begin{equation}
\{\psi({\bf x}),\psi^\dagger({\bf y})\}=\delta({\bf x-y}), \ \ \{\psi({\bf
x}),\psi({\bf y})\}=0,
\end{equation} and also that,
\begin{equation}
\{\psi_{\mbox{\scriptsize GI}}({\bf x}),\psi^\dagger_{\mbox{\scriptsize
GI}}({\bf
y})\}=\delta({\bf x-y}),
\ \
\{\psi_{\mbox{\scriptsize GI}}({\bf x}),\psi_{\mbox{\scriptsize GI}}({\bf
y})\}=0.
\end{equation} The gauge-invariant `electrons' and `positrons' in the
transformed $\,\hat{}\,$ representation therefore are anticommuting
fermions  that
interact through the nonlocal interactions $H_{\mbox{\scriptsize a}}$ and
$H_{\mbox{\scriptsize b}}$,  and through the photon-current interaction
$H_{j\gamma}$.\bigskip

\noindent{\begin{underline}{Rotational phases can be arbitrary in this
model:}\end{underline}}\\ Consider the rotation operator in the transformed
$\hat{\rule{6pt}{0pt}}$ representation.
$\hat{J} = J + {\cal J}$ where
\begin{eqnarray}
 J &=& -\int d{\bf x}\ \Pi_lx_i\epsilon_{ij}\partial_jA_l + \int d{\bf x}\
Gx_i\epsilon_{ij}\partial_jA_0 - \int d{\bf x}\
\epsilon_{ij}\Pi_iA_j\nonumber\\
&-&i\int
d{\bf x}\ \psi^\dagger x_i\epsilon_{ij}\partial_j\psi -
\frac{1}{2}\,\int d{\bf x}\ \psi^\dagger\gamma_0\psi
\end{eqnarray} and
\begin{eqnarray} {\cal J} &=& -\sum_{\bf
k}\frac{m^{3/2}}{16k^3}\,\epsilon_{ln}k_l\frac{\partial\phi({\bf k})}{\partial
k_n}\,j_0(-{\bf k})j_0({\bf k})\nonumber\\ &-& \sum_{\bf k}
i\epsilon_{ln}k_l\frac{\partial\theta({\bf k})}{\partial k_n}\left[a_Q({\bf
k})j_0(-{\bf k}) - a_Q^{\mbox{\normalsize $\star$}}({\bf k})j_0({\bf
k})\right]\nonumber\\ &+& \sum_{\bf k}\epsilon_{ln}k_l\frac{\partial\phi({\bf
k})}{\partial k_n}\left[a_Q({\bf k})j_0(-{\bf k}) + a_Q^{\mbox{\normalsize
$\star$}}({\bf k})j_0({\bf k})\right]
\end{eqnarray} If $\theta(k)=\phi(k) = 0$ (a perfectly consistent and viable
choice), ${\cal J} = 0,$ $\hat{J} = J,$  and the charged states in the
transformed $\hat{\rule{6pt}{0pt}}$ representation rotate {\em precisely} as in
the untransformed representation.
 But we can also choose, for example,
\begin{equation}
\phi({\bf k}) = -\frac{8k^2}{m^{5/2}}\,\delta(k)\,\tan^{-1}\frac{k_2}{k_1},
\end{equation} for which ${\cal J} = Q^2/4\pi m$ ($Q$ is the charge operator)
and, under a $2\pi$ rotation, an electron state  in the transformed
$\hat{\rule{6pt}{0pt}}$ representation picks up the arbitrary ``anyonic''  phase
$\displaystyle e^{i(e^2/4\pi m)}$. {\em This does not affect the statistical
behavior, however.}\bigskip

\noindent{\begin{underline}{Particle exchange:}\end{underline}}\\ Consider a
state
$|A\rangle=e^\dagger({\bf k})e^\dagger(-{\bf k})|0\rangle$. If we use
$e^{i(J+Q^2/4\pi m)\pi}$ to rotate $|A\rangle$ through $\pi$ so that ${\bf k}
\rightleftharpoons -{\bf k}$, we get
\begin{equation} |A^\prime\rangle = -e^{ie^2/m}e^\dagger(-{\bf k})e^\dagger({\bf
k})|0\rangle
\end{equation} Suppose we interchange $e^\dagger({\bf k})$ and $e^\dagger(-{\bf
k})$ by commuting them. For the  purposes of this discussion, we allow for the
possibility of an `exotic'  graded commutator algebra:
\begin{equation} e^\dagger({\bf k})e^\dagger(-{\bf
k})+e^{i\delta}e^\dagger(-{\bf
k})e^\dagger({\bf k})=0.
\end{equation} If we exchange $e^\dagger({\bf k})$ and $e^\dagger(-{\bf k})$ in
$|A^\prime\rangle$ with that commutator algebra,  we get
\begin{equation} |A^{\prime\prime}\rangle = e^{i(e^2/m+\delta)}e^\dagger(-{\bf
k})e^\dagger({\bf k})|0\rangle
\end{equation} Can we argue that the phase factor $e^{i(e^2/m+\delta)}$ must
equal 1, so that the  statistical properties of this model must match the
arbitrary rotational phase? Our conclusion is that such an argument can not be
supported.

In $3+1$ dimensions, the Lorentz group is so constraining that arbitrary
rotational phases are not allowed.  Vacuum expectation values of
Lorentz-transformed fields are unique. Adding constant (like $Q^2/4\pi m$) to
${\bf J}$ in $3+1$ dimensions would violate $[J_i, J_j] =  i\epsilon_{ijk}J_k$.
The same feature that allows arbitrary rotational phases in $2+1$ dimensions,
also affects the applicability, to this  (2+1)-dimensional model, of the
`standard'
proof of the spin-statistics connection.\cite{burgy,sw,lzumino} One of the
relatively few assumptions that underlie these proofs is the structure of the
(complex) Lorentz group, which, in $3+1$ dimensions, uniquely associates a $-1$
factor with a $2\pi$ rotation of a Dirac spinor. That makes it necessary to
reexamine the spin-statistics connection in spaces that allow arbitrary
rotational
phases.

\acknowledgements
This research was supported by the Department of Energy
under Grant No. DE-FG02-92ER40716.00.

\end{document}